\documentclass[reprint,amsmath,amssymb]{elsarticle}

\usepackage{graphicx}% Include figure files
\usepackage{subfig}
\graphicspath{{./Immagini/}}
\usepackage{dcolumn}% Align table columns on decimal point
\usepackage{bm}% bold math
\usepackage{hyperref}% add hypertext capabilities
\usepackage[mathlines]{lineno}% Enable numbering of text and display math
\usepackage[export]{adjustbox}
\usepackage{color}
\usepackage{tabularx,array,booktabs}
\usepackage{caption}
\usepackage{mathtools}
\usepackage{amssymb}
\usepackage[normalem]{ulem}
\usepackage{comment}

\newcommand{\Om}{\Omega}
\newcommand{\om}{\omega}
\newcommand{\X}{\boldsymbol{X}}
\newcommand{\x}{\boldsymbol{x}}

\newcommand{\F}{\boldsymbol{F}}
\newcommand{\V}{\boldsymbol{V}}
\newcommand{\R}{\boldsymbol{R}}

\newcommand{\editato}[1]{\textcolor{black}{#1}} %blue

\begin{document}

%\preprint{APS/123-QED}
\begin{frontmatter}

\title{Experimental validation of a broadband pentamode elliptical-shaped cloak for underwater acoustics}

\author[polimi]{Davide Enrico Quadrelli\fnref{myfootnote}}
\author[polimi]{Matteo Alessandro Casieri}
\author[polimi]{Gabriele Cazzulani}
\author[marina]{Simone La Riviera}
\author[polimi]{Francesco Braghin}
\fntext[myfootnote]{davideenrico.quadrelli@polimi.it}

\address[polimi]{Department of Mechanical Engineering, Politecnico di Milano, via La Masa 1, 20156 Milano, Italy}
\address[marina]{Stato Maggiore della Marina, 5$^o$ Reparto Sommergibili, Piazza della Marina 4, 00196 Roma}
\date{\today}

\begin{abstract}
In this paper we report the numerical design and experimental validation of a pentamode cloak for scattering reduction of elliptical targets in underwater acoustics. Despite being a well-known theoretical concept, experimental validations of transformation-based acoustic cloaking for non-axisymmetric targets have been hindered due to the difficulties related to the complexity of the required material distribution. We overcome these by adopting a linear quasi-symmetric map defined in elliptic coordinates that allows for the design of a pentamode cloak with constant anisotropic elasticity and a scalar inhomogeneous density. We then perform the long-wavelength homogenization based optimization of unit cells that implement the computed material distribution. \editato{Numerical validations allow to assess the working range of the designed microstructure, which depends solely on the ratio between the wavelength and the size of the microstructure, and is thus expressed in terms of the maximum normalized wavenumber $\kappa L_c$, being $L_c$ the characteristic length of the considered unit cell. The influence on the scattering performance of the cloak of a pentamode bandgap common to all the cells in the microstructure is also discussed, finding that it is actually detrimental for the scattering reduction. Finally, the specimen is fabricated and experimentally tested over the range $0.45<\kappa L_c<1.34$.}
\end{abstract}


\end{frontmatter}
%\maketitle

%%%%%%%%%%%%%%%%%%%%%%%%%%%%%%%%%%%%%%%%%%%%%%%%%%%%%%%%%%%%%%%%%%%%%%%%%
%%%%%%%%%%%%%%%%%%%%%%%%%%%%%%%%%%%%%%%%%%%%%%%%%%%%%%%%%%%%%%%%%%%%%%%%
\section{Introduction}
%%%%%%%%%%%%%%%%%%%%%%%%%%%%%%%%%%%%%%%%%%%%%%%%%%%%%%%%%%%%%%%
%%%%%%%%%%%%%%%%%%%%%%%%%%%%%%%%%%%%%%%%%%%%%%%%%%%%%%%%%%%%%%%
The possibility of reducing the acoustic signature of underwater targets has generated considerable interest on acoustic cloaking \cite{chen2010acoustic} during the past decade. Firstly inspired by the findings on transformation theory in electromagnetics \cite{Pendry2006, Leonhardt2006}, this concept was lately formalized as a discipline in the acoustic field by Norris \cite{Norris2008}, who pointed out the non-uniqueness of the material distribution that realizes the invisibility cloak for a given transformation.

Acoustic cloaks can be made, for example, by fluids characterized by anisotropic inertial properties and scalar bulk modulus. This type of solution, called \textit{inertial cloak} (IC) \cite{cummer2007one, chen2007acoustic} can physically be implemented either with layered fluids \cite{torrent2008acoustic}, or with cavities filled by the hosting background fluid, adopting the transmission line analogy. This latter approach, allowed Fang and coworkers \cite{zhang2011broadband} to provide the first experimental validation of a broadband underwater acoustic cloak, designed to render acoustically invisible an axisymmetric target.

Conversely, the required anisotropy for cloaking can be also realized adopting \textit{pentamode materials}, a special class of solids characterized by singular anisotropic elasticity tensors \cite{milton1995elasticity}. Their solid nature, the bounded density and the possibility to avoid working fluids have made pure \textit{pentamode cloaks} (PM) with scalar density a promising alternative for practical implementation. Indeed, Chen et al. \cite{chen2015latticed, chen2017broadband} have recently designed and experimentally validated an axisymmetric layered cloak based on homogenization driven optimization of pentamode microstructures. 

Nonetheless, no report of experimental validations of cloaking based on transformation acoustics has been documented for targets with shapes different from the axisymmetrical cylinder. Li et al. \cite{li2012homogeneous} and Vipperman and coworkers \cite{li2014two, li2018non, li2019two}  have firstly tackled the problem of designing acoustic cloaks for arbitrarily shaped targets, showing a design method based on the IC concept. On the other side of possible material distributions, Chen at al. \cite{chen2016design} proposed a numerical approach to retrieve quasi-symmetric transformations for the design of pure PM for arbitrary shapes. The necessary and sufficient condition for designing PM cloaks with isotropic inertia is indeed satisfied when the transformation gradient is a pure stretch \cite{Norris2008}.

In this paper we instead follow a different path to design a PM acoustic cloak for elliptic targets by using an analytical quasi-symmetric linear map defined in elliptic coordinates. It has been shown \cite{quadrelli2021acoustic} that this method allows for obtaining near-cloaks comprising anisotropic homogeneous elasticity tensors and isotropic inhomogeneous density. The introduced approximation due to the local rotation being non zero can be quantified and shown to be small for the chosen configuration. The resulting homogeneous elastic properties of the designed cloak are crucial for simplification of the optimization of the microstructure, since a common elastic frame implementing the required anisotropic stiffness can in that case be repeated almost unchanged everywhere in the cloak. The density gradient is then realized adopting localized masses that have little impact on the homogenized elastic tensor but allow for local variation of the density. This simplification allows to keep the computational burden of the optimization as low as possible: the lack of axial symmetry, indeed, prevents the possibility of discretizing continuous distributions in layers with constant properties as was done in previous works \cite{zhang2011broadband, chen2017broadband}. This fact renders the total number of different unit cells to be designed one order of magnitude higher than in previous experimental realizations, where material properties vary only in the radial direction. Adopting a map in elliptic coordinates allows also to obtain principal directions of anisotropy everywhere tangential and normal to confocal ellipses, which simplifies the assembly phase of the optimized unit cells as a whole microstructure, avoiding irregular orientations that could be cumbersome to implement. \editato{Numerical and experimental tests allow to shed light on the performance of the designed microstructure. In particular, the influence of a pentamode bandgap appearing in the same frequency range for all the designed unit cells is discussed. It could be speculated, indeed, that the single mode bandgap is the actual working frequency range for a microstructured pentamode cloak, since no bulk propagating shear waves can be supported by the cloak, thus leading to the required fluid-like behavior. Our numerical findings, however, show that from the perspective of the scattering cross section of the cloak, the pentamode bandgap witness the worst of the performance. The working range of the cloak is thus solely related to the long-wavelength regime, and thus the normalized wavenumber $\kappa L_c$, being $L_c$ a characteristic dimension of the microstructure, is used to characterize the scattering response of the system. Experimental tests allow to discuss the impact of absorption on the actual behavior of the cloak.} 

The manuscript is organized as follows: after this brief introduction, the theoretical framework related to transformations in elliptic coordinates is recalled, and the configuration chosen for the experimental validation is introduced. In the third section the design of the microstructure is analyzed, starting by the optimization of the unit cells and their conformal assembly around the target, to conclude with the numerical validation of the cloak. The fourth section deals with the apparatus and the instrumentation adopted to retrieve the experimental results. Before drawing conclusions, the fifth section is devoted to the discussion of experimental results.

%%%%%%%%%%%%%%%%%%%%%%%%%%%%%%%%%%%%%%%%%%%%%%%%%%%%%%%%%%%%%%%%%%%%
%%%%%%%%%%%%%%%%%%%%%%%%%%%%%%%%%%%%%%%%%%%%%%%%%%%%%%%%%%%%%%%%%%%%
\section{Quasi-symmetric transformation}
%%%%%%%%%%%%%%%%%%%%%%%%%%%%%%%%%%%%%%%%%%%%%%%%%%%%%%%%%%%%%%%
%%%%%%%%%%%%%%%%%%%%%%%%%%%%%%%%%%%%%%%%%%%%%%%%%%%%%%%%%%%%%%%
%%%%%%%%%%%%%%%%%%%%%%%%%%%%%%%%%%%%%%%
%%%%%%%%%%%%%%%%%%%%%%%%%%%%%%%%%%%%%%%
\begin{figure*}[ht]
\begin{center}
\includegraphics[width = .8\textwidth]{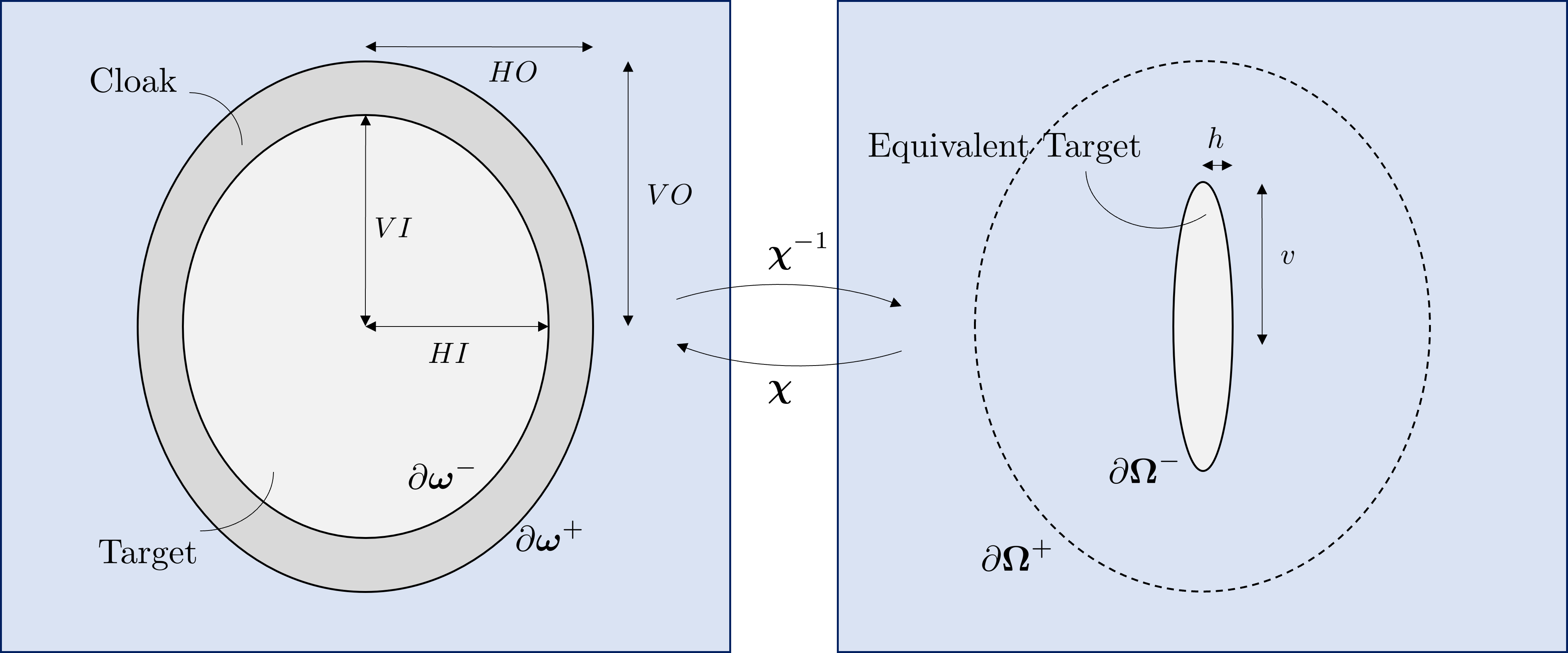}
\caption{\label{fig:FIG1}{Schematic representation of the geometrical features of the cloak and the map between undeformed virtual domain and physical space.}}
\end{center}
\end{figure*}
%%%%%%%%%%%%%%%%%%%%%%%%%%%%%%%%%%%%%%%
%%%%%%%%%%%%%%%%%%%%%%%%%%%%%%%%%%%%%%%
Adopting the language of finite deformations, we consider a map $\x = \boldsymbol{\chi}( \X)$ between points in the undeformed virtual domain $\boldsymbol{\Om}$ and points in the deformed domain $\boldsymbol{\om}$ (Figure \ref{fig:FIG1}), that can be represented in elliptic coordinates as \cite{quadrelli2021acoustic}:
\begin{equation}
\Om = \left\{ \X : (X, Y)=(a \sinh(R) \sin \Theta, a \cosh(R) \cos\Theta),  R \in [R_1,R_3], \Theta \in [0,2\pi] \right\}
\end{equation}
and
\begin{equation}
\om = \left\{ \x : (x,y) =(a \sinh(r) \sin \theta, a \cosh(r)\cos \theta), r \in [R_2,R_3], \theta \in [0,2\pi] \right\}
\end{equation}
A given constant $r$ or $R$ value describes confocal ellipses, while constant $\theta$ or $\Theta$ describe confocal hyperbolae. Capital letters are used for the undeformed configuration, while plain letters for the deformed one. The constant quantity $a$ represents the location of the foci on the vertical axis. Points characterized by $r=R_2$ and $r,R=R_3$ represent points on the inner boundary and outer boundaries of the cloak, while points characterized by $R=R_1$ lie on the surface $\partial \boldsymbol{\Omega}^-$ to which the inner boundary $\partial \boldsymbol{\om}^-$ is mapped by the inverse transformation, and thus represent the "equivalent" acoustic behavior of the near-cloak. The geometrical features of these surfaces can be equivalently expressed via the vertical and horizontal semiaxes, whose length can be obtained as:
\begin{equation}
    VO=a \cosh(R_3); \qquad HO=a \sinh(R_3)
\end{equation}
for the outer surface $\partial \boldsymbol{\om}^+,\;\partial \boldsymbol{\Om}^+$, as depicted in Figure \ref{fig:FIG1}. Similarly, the semiaxes $VI$ and $HI$ for the inner surface and $v$, $h$ for $\partial \boldsymbol{\Omega}^-$ can be obtained using $R_2$ and $R_1$ respectively.
A generic inverse map can be written as:
\begin{equation}
\begin{cases}
R=f(r)\\
\Theta=\theta
\end{cases}
\end{equation}
which leads to the following deformation gradient:
\begin{equation}
    \F=\V\R=(\lambda_1\boldsymbol{n}_1 \otimes \boldsymbol{n}_1+\lambda_2 \boldsymbol{n} \otimes \boldsymbol{n}_2)(\boldsymbol{n}_1 \otimes \boldsymbol{N}_1+ \boldsymbol{n}_2\otimes \boldsymbol{N}_2)=\lambda_1\boldsymbol{n}_1\otimes\boldsymbol{N}_1+\lambda_2\boldsymbol{n}_2\otimes\boldsymbol{N}_2
    \label{eq_Fgeneral}
\end{equation}
being
\begin{equation}
\begin{split}
&\lambda_1=\frac{1}{f^\prime(r)}\frac{\sqrt{\sinh^2(r)+\sin^2 \theta}}{\sqrt{\sinh^2(f(r))+\sin^2 \theta}}\\
&\lambda_2=\frac{\sqrt{\sinh^2(r)+\sin^2 \theta}}{\sqrt{\sinh^2(f(r))+\sin^2 \theta}}
\end{split}	
\end{equation}
the principal stretches, and $\boldsymbol{n}_i$,  $\boldsymbol{N}_i$ unit vectors tangent and normal to the coordinate ellipse at the considered point.
Being, in general, $\boldsymbol{n}_i\neq \boldsymbol{N}_i$ (without summation implied) the transformation is not a pure stretch. It has been shown \cite{quadrelli2021acoustic} that in most feasible situations the rotation remains bounded to low values all over the cloak. If such condition is met then the material properties, normalized with respect to those of the background fluid, are:
\begin{equation}
\begin{split}
&\rho=J^{-1}=(\lambda_1\lambda_2)^{-1}=f^\prime(r)\frac{\sinh^2(f(r))+\sin^2 \theta}{\sinh^2(r)+\sin^2 \theta}\\
&K=J=\lambda_1\lambda_2=\frac{1}{f^\prime(r)}\frac{\sinh^2(r)+\sin^2 \theta}{\sinh^2(f(r))+\sin^2 \theta}\\
&\boldsymbol{S}=J^{-1}\V=\frac{\sqrt{\sinh^2(f(r))+\sin^2 \theta}}{\sqrt{\sinh^2(r)+\sin^2 \theta}}\boldsymbol{n}_1 \otimes \boldsymbol{n}_1 + f^\prime(r)\frac{\sqrt{\sinh^2(f(r))+\sin^2 \theta}}{\sqrt{\sinh^2(r)+\sin^2 \theta}} \boldsymbol{n}_2 \otimes \boldsymbol{n}_2\\
&\mathbb{C}=K \boldsymbol{S} \otimes \boldsymbol{S} 
\end{split}
\end{equation}
where $\rho$ is the density distribution, $\mathbb{C}$ the fourth order singular elasticity tensor, $K$ and $\boldsymbol{S}$ are its non-zero eigenvalue and associated  eigenvector, respectively.
For the experimental implementation of the cloak we choose a linear map:
\begin{equation}
    f(r)=\frac{R_3-R_1}{R_3-R_2}(r-R_2)+R_1
\end{equation}
that results in the following constant elasticity tensor, written in Voigt notation:
\begin{equation}
[C]=K_0\begin{bmatrix}
    \displaystyle\frac{1}{f^\prime(r)} & 1 & 0 \\
    1 & f^{\prime}(r) & 0\\
    0 & 0 & 0
    \end{bmatrix}=
    K_0\begin{bmatrix}
    \displaystyle\frac{R_3-R_2}{R_3-R_1} & 1 & 0 \\
    1 & \displaystyle\frac{R_3-R_1}{R_3-R_2} & 0\\
    0 & 0 & 0
    \end{bmatrix}=\begin{bmatrix}
    K_r & K_{r\theta} & 0 \\
    K_{r\theta} & K_\theta & 0\\
    0 & 0 & 0
    \end{bmatrix}
\end{equation}
where $K_0$ is the bulk modulus of the background fluid, and $K_r$, $K_\theta$, $K_{r \theta}$ are introduced to indicate the entries of the elasticity matrix.
\editato{Once the shape of the target is assigned,} the only two parameters that are left to be chosen are $R_1$ and $R_3$. These set, respectively, the shape of the boundary $\partial \boldsymbol{\Omega}^-$ and the shape of $\partial \boldsymbol{\om}^+$. 
%%%%%%%%%%%%%%%%%%%%%%%%%%%%%%%%%%%%%%%
%%%%%%%%%%%%%%%%%%%%%%%%%%%%%%%%%%%%%%%
\begin{figure*}[t]
\begin{center}
\includegraphics[width = \textwidth]{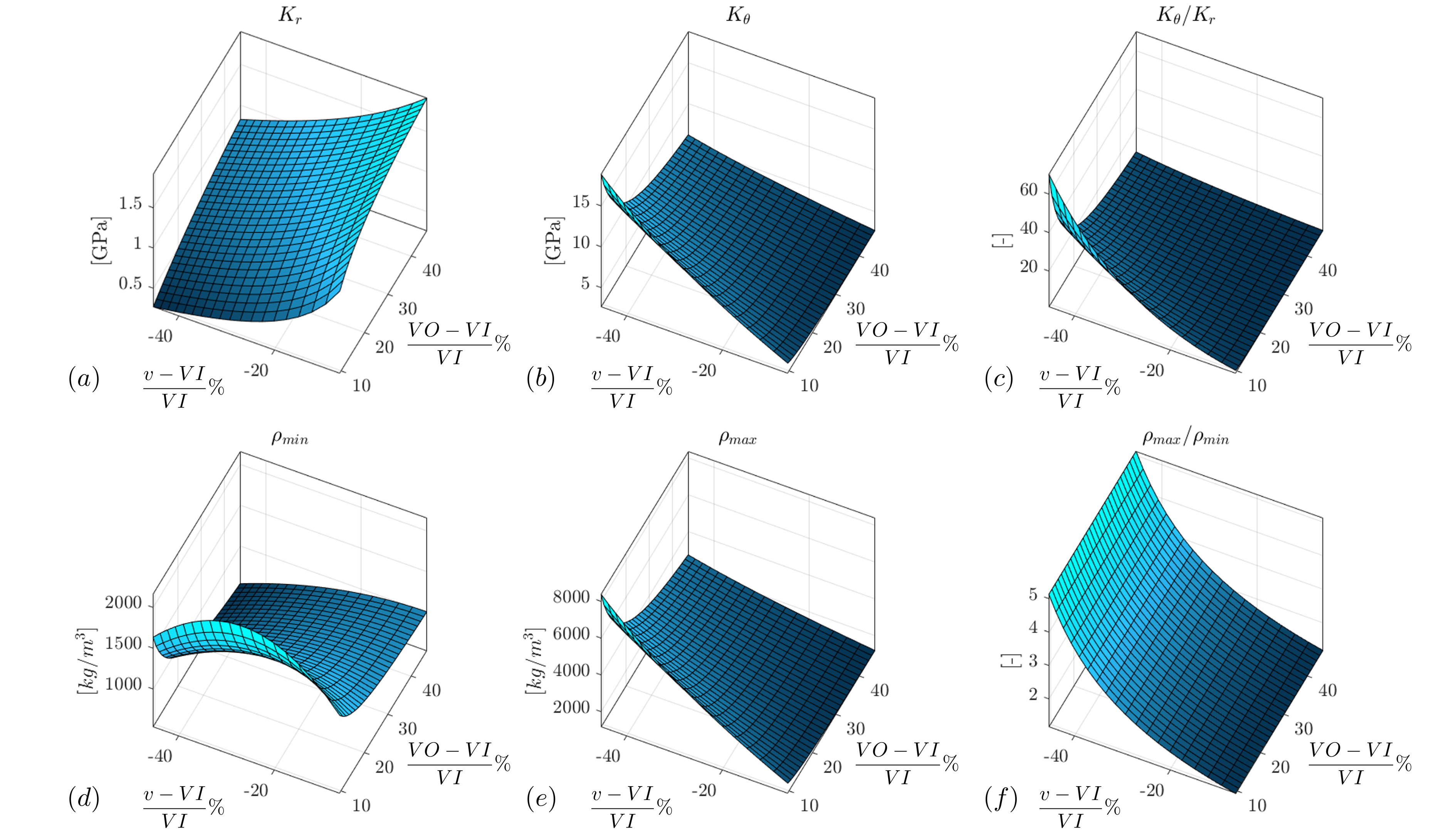}
\caption{\label{fig:FIG2}{\ref{fig:FIG2}(a): Variation of the bulk modulus $K_r$ as a function of the geometrical features of the cloak expressed as equivalent percentage reduction in vertical semiaxis $(v-VI)/VI \%$ and percentage increase in vertical semiaxis $(VO-VI/VI \%)$. \ref{fig:FIG2}(b): Map for the bulk modulus $K_\theta$. \ref{fig:FIG2}(c): Anisotropy factor $K_\theta/K_r$. \ref{fig:FIG2}(d): Minimum density to be implemented in the cloak $\rho_{min}$. \ref{fig:FIG2}(e): Maximum density $\rho_{max}$. \ref{fig:FIG2}(f): Inhomogeneity factor $\rho_{max}/\rho_{min}$.}}
\end{center}
\end{figure*}
%%%%%%%%%%%%%%%%%%%%%%%%%%%%%%%%%%%%%%%
%%%%%%%%%%%%%%%%%%%%%%%%%%%%%%%%%%%%%%%
These parameters can be chosen adopting the maps represented in Figure \ref{fig:FIG2}, where the variability of material properties with the variation of the design parameters is considered. To get a more appreciable physical insight, the variation of the geometry of the cloak is expressed through the percentage variation in "thickness" $(VO-VI)/VI$, defined by $R_3$, and through the equivalent major semiaxis reduction $(v-VI)/VI$, related to $R_1$. Indeed, due to the nature of the adopted transformation, the target cannot be shrunk to a point, but in the limit with $R_1 \rightarrow 0$ the outer surface of the target is mapped to the vertical line joining the two foci \cite{quadrelli2021acoustic}. For this reason, the performance will be dependent on the direction of incidence: vertical incidence, when plane waves travel in the direction parallel to $v$, is the best case scenario, while horizontal incidence, conversely, is the worst case. It follows that the aim is not to render completely invisible the target, but to obtain a reduction in its scattering cross section that is appreciable for practical applications. Fabrication of the specimen in a single piece implies that all the unit cells realizing the inhomogeneous and anisotropic material distribution must be made by the same material, which is chosen to be Aluminum ($\rho=2700\;[kg/m^3]$, $E=70\;[GPa]$, $\nu=0.3$). This sets the bounds on the maximum and minimum values of homogenized material properties that can be obtained, the bottleneck being the range of densities that can be obtained varying the filling fraction of a single material. As a consequence, this sets also the bounds on the minimum value of $R_1$, i.e. $(v-VI)/VI$, whose decrease causes an increase of the ratio between the maximum and minimum density in the cloak (Figure \ref{fig:FIG2}(f)), and on the minimum $R_3$, i.e. $(VO-VI)/VI$, whose decrease causes an increase both in the minimum and maximum values of density (Figure \ref{fig:FIG2}(d) and \ref{fig:FIG2}(e)). Looking at the maps in Figure \ref{fig:FIG2}, the values $R_1=7.8/12R_2$ and $R_3=1.244R_2$ are chosen, which correspond to:
\begin{equation}
    \frac{v-VI}{VI}\approx-40\%; \qquad \frac{VO-VI}{VI}\approx47\%.
\end{equation}
Lower reduction in the thickness of the cloak can be obtained using a more dense material like Titanium, and lower values of $v$ can be reached by using different materials in different locations of the cloak to get an overall wider range of available densities.
%%%%%%%%%%%%%%%%%%%%%%%%%%%%%%%%%%%%%%%%%%%%%%%%%%%%%%%%%%%%%%%
%%%%%%%%%%%%%%%%%%%%%%%%%%%%%%%%%%%%%%%%%%%%%%%%%%%%%%%%%%%%%%%
\section{Unit cell design and numerical validation}
%%%%%%%%%%%%%%%%%%%%%%%%%%%%%%%%%%%%%%%%%%%%%%%%%%%%%%%%%%%%%%%
%%%%%%%%%%%%%%%%%%%%%%%%%%%%%%%%%%%%%%%%%%%%%%%%%%%%%%%%%%%%%%%
%%%%%%%%%%%%%%%%%%%%%%%%%%%%%%%%%%%%%%%
%%%%%%%%%%%%%%%%%%%%%%%%%%%%%%%%%%%%%%%
\begin{figure*}[th!]
\begin{center}
\includegraphics[width = \textwidth]{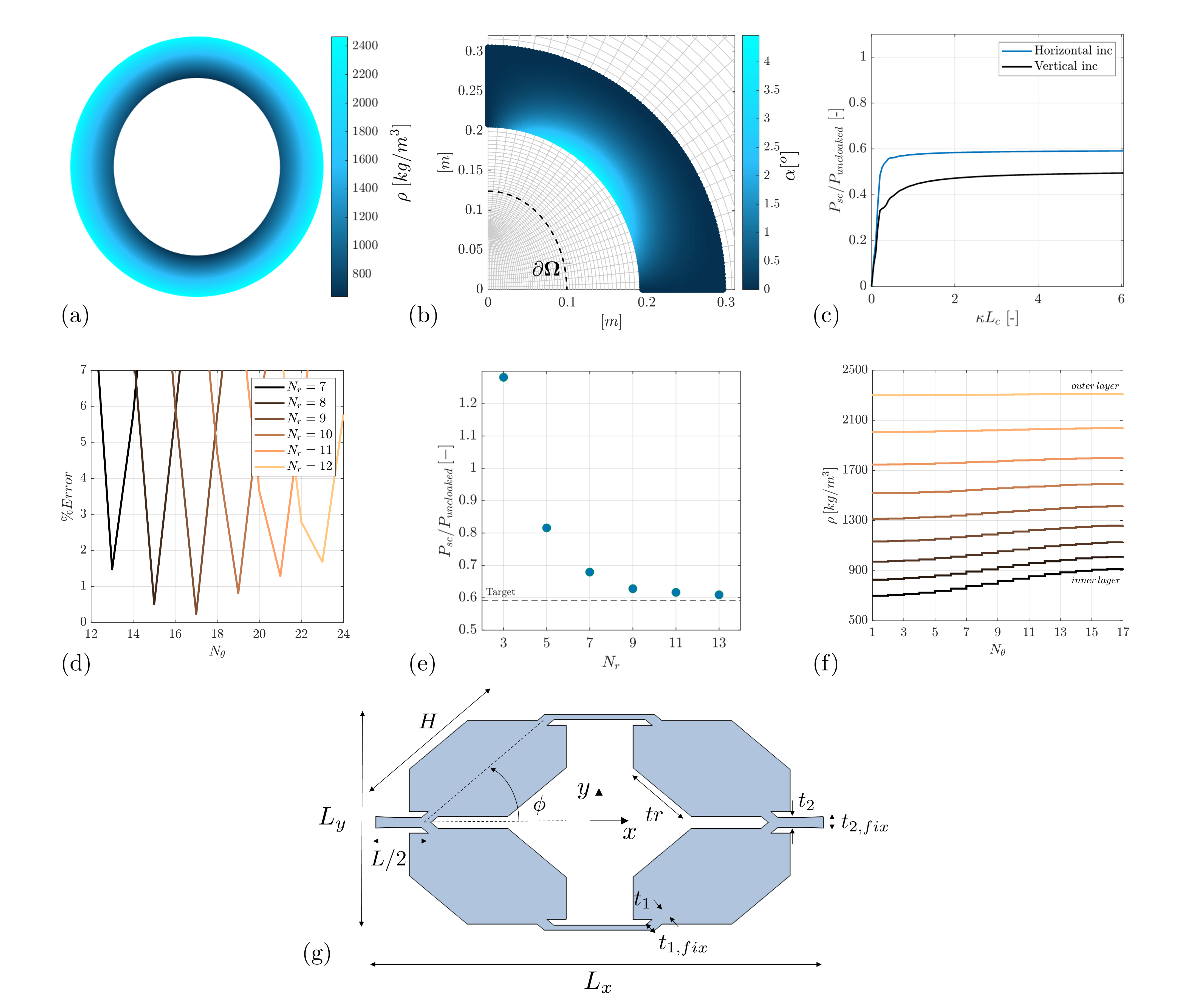}
\caption{\label{fig:FIG3}{\ref{fig:FIG3}(a): Density distribution in the cloak. \ref{fig:FIG3}(b): Local rotation angle for the linear map. Coordinate lines are drawn in gray, while the boundary $\partial \boldsymbol{\Omega}^-$ is represented with a dashed black line. \ref{fig:FIG3}(c): Acoustic performance computed as the ratio of scattered power between the equivalent target $\partial \boldsymbol{\Omega}^-$ and the uncloaked case, for vertical and horizontal incidence. \ref{fig:FIG3}(d): Percentage error on the conformal map for the discretization of the continuous density distribution, as a function of the assigned number of radial elements for varying the number of tangential elements \ref{fig:FIG3}(e): Performance of the numerical model built discretizing the continuous density distribution, for different numbers of radial elements. \ref{fig:FIG3}(f): Discretized density distribution with $N_r=9$ and $N_\theta=17$. \ref{fig:FIG3}(g): Schematic view of the geometry of the pentamode unit cell and the parameters adopted for the optimization of homogenized properties.}}
\end{center}
\end{figure*}
%%%%%%%%%%%%%%%%%%%%%%%%%%%%%%%%%%%%%%%
%%%%%%%%%%%%%%%%%%%%%%%%%%%%%%%%%%%%%%%
Figure \ref{fig:FIG3}(a) represents the computed density distribution for the chosen configuration, along with the computed local rotation $\alpha$ associated to the lack of perfect symmetry of the deformation tensor (Figure \ref{fig:FIG3}(b)). Since the rotation is bounded to low values across the entire surface of the cloak, the acoustic scattering from the cloaked obstacle can be considered equal to that of the boundary $\partial \boldsymbol{\Omega}^-$ \cite{quadrelli2021acoustic}. The performance of the cloak can be thus inspected by computing the ratio of the scattered power $P_{sc}$ of the equivalent target $\partial \boldsymbol{\Om}^-$ and the scattered power in the uncloaked scenario $P_{uncloaked}$, for plane wave incidence. The scattered power is computed adopting the finite element method implemented in the commercial software COMSOL Multiphysics, by integrating the scattered intensity from acoustically rigid obstacles having the appropriate elliptic geometry. The results are shown in Figure \ref{fig:FIG3}(c), for vertical (best case scenario) and horizontal (worst case scenario) direction of incidence. \editato{For ease of comparison with the following numerical results computed on the microstructure, the frequency dependence is shown in terms of the wavenumber normalized with respect of the characteristic length of the microstructure $L_c$ that is introduced below.} Note that in the geometrical acoustics limit the total scattering cross section of a rigid obstacle tends to two times its geometric cross section, thus the ratio of scattered power $P_{sc}/P_{uncloaked}$ at high frequencies tends to the ratio between the semiaxes $v/VI$ and $h/HI$ for horizontal and vertical incidence, respectively.  

The computed target pentamode elasticity tensor is:
\begin{equation}
    [C]=\begin{bmatrix}
    K_{r}& K_{r\theta} & 0\\
    K_{r\theta} & K_{\theta} & 0\\
    0   &  0 & 0
    \end{bmatrix}=\begin{bmatrix}
    0.91 & 2.25 & 0\\
    2.25 & 5.55 & 0\\
    0   &  0 & 0
    \end{bmatrix}\; [GPa]
    \label{ElastTarget}
\end{equation}
%%%%%%%%%%%%%%%%%%%%%%%%%%%%%%%%%%%%%%%
%%%%%%%%%%%%%%%%%%%%%%%%%%%%%%%%%%%%%%%
%\begin{figure*}[t]
%\begin{center}
%\includegraphics[width = 0.8\textwidth]{Immagini/Fig4bis.png}
%\caption{\label{fig:FIG4}{Schematic view of the geometry of the pentamode unit cell and the parameters adopted for the optimization of homogenized properties.}}
%\end{center}
%\end{figure*}
%%%%%%%%%%%%%%%%%%%%%%%%%%%%%%%%%%%%%%%
%%%%%%%%%%%%%%%%%%%%%%%%%%%%%%%%%%%%%%%
The required material properties are obtained in practice by optimization of the long wavelength slopes of the band structure of the unit cell depicted in Figure \ref{fig:FIG3}(g), by imposing:
\begin{equation}
    \begin{split}
        &C_{yy}=c_{Ly}^2\rho=K_r\\
        &C_{xx}=c_{Lx}^2\rho=K_\theta\\
        &C_{xy}=\sqrt{\rho^2(c_{qL}^2-c_{qS}^2)^2-(C_{xx}-C_{yy})^2/4}-G=K_{r \theta}\\
        &G=c_{S}^2\rho \rightarrow 0
    \end{split}
\end{equation} 
It is implied that the cell will be accommodated around the target in such a way that its $x$ direction (Figure \ref{fig:FIG3}(g)) is aligned to the direction tangential to coordinate ellipses and the $y$ direction to the normal direction of coordinate ellipses. Indeed, $c_{Lx}$ and $c_{Ly}$ represent the wave speed of the longitudinal modes computed in the linear part of the dispersion relation for propagation in the $x$ and $y$ directions, respectively, while $c_{qL}$ and $c_{qS}$ are the speeds of quasi-longitudinal and quasi-shear modes computed for propagation along the $45^o$ direction. The speed of the shear mode in $x$ direction is indicated by $c_S$. The homogenized density $\rho$ is computed as the density of Aluminum multiplied by the filling fraction, since no resonance mechanism is observed. 
The topology of the unit cell is inspired by the one adopted by Chen et al. in \cite{chen2015latticed, chen2017broadband}. It is made by a hexagonal frame that is ideally responsible of the homogenized elastic behavior of the unit cell, while big appendages are included to control the equivalent density in the most uncoupled way as possible with respect to the equivalent stiffness. The elasticity tensor being constant, indeed, allows for a preliminary optimization aimed at obtaining the geometry of the hexagonal frame that exhibits the needed elastic properties, that is next maintained as constant as possible across the overall cloak. This reduces the computational effort and at the same time facilitates connectivity between different adjacent cells. The first step in the microstructure design is thus to set a GA-based optimization of a single unit cell targeting the elasticity tensor in Equation \ref{ElastTarget}. This is done adopting Matlab optimization routines in conjunction with Bloch analysis of the unit cell performed with Comsol Multiphysics. In this preliminary optimization the only parameters considered as design variables are $\phi$ and $L/H$, while all other parameters are set manually before optimization runs in order to get close to a homogenized density in between the maximum and minimum value inside the cloak. The cost function adopted at this stage is:
\begin{equation}
    J=\frac{|C_{xx}-K_{\theta }|}{K_{\theta}}+\frac{|C_{yy}-K_{r}|}{K_{r}}
\end{equation}
The other elements of the elastic tensor are not included in the cost function, since it can be verified that when the size of the junctions between beams is kept low, $G$ naturally tends to values two orders of magnitude smaller than $K_{\theta}$ and $C_{xy}\rightarrow \sqrt{K_{r}K_{\theta}}$. For this reason and to ease connectivity between cells, the values of $t_{1,fix}$ and $t_{2,fix}$ are kept fixed for all the stages of optimization. In the same way, $\phi$ and $L/H$ are from considered to be constant after this stage. These parameters set the geometry of the hexagonal frame that is kept equal for the whole microstructure of the cloak and, in particular, univocally determine the aspect ratio $L_x/L_y$ of the rectangular unit cell. This parameter is important since it is linked to the way in which the continuous density distribution must be discretized for practical implementation. Indeed, the discretization should be done in such a way to allow accommodation of the microstructure around the target without distortions in its shape, that could introduce unwanted anisotropy. The elliptic coordinate system, being a orthogonal system with equal scaling factors, allows for definition of conformal maps in an easy way by linearly mapping the cartesian coordinates to elliptical ones \cite{quadrelli2021acoustic}. Indeed, due to the equal scaling factors, the aspect ratio between blocks obtained discretizing the domain $\boldsymbol{\om}$ using increments $dr$ and $d\theta$ in elliptic coordinates results in blocks having aspect ratio:
\begin{equation}
    \frac{L_\theta}{L_r}=\frac{d\theta}{dr}
\end{equation}
This aspect ratio should be the same as the previously found $L_x/L_y$ for the unit cell in cartesian coordinates, which means that the number of cells $N_r$ and $N_\theta$ in "radial" and "tangential" directions over one quarter of cloak can't be chosen independently but should satisfy:
\begin{equation}
    \frac{d\theta}{dr}=\frac{\pi/2}{N_\theta}\frac{N_r}{R_3-R_2}=\frac{L_x}{L_y}
\end{equation}
Considering that $N_r$ and $N_\theta$ can take only integer values, the exact solution may not exist, thus the percentage error
\begin{equation}
    e_{\%}= \left( \frac{L_\theta/L_r-L_x/L_y}{L_x/L_y} \right) \cdot 100
\end{equation}
should be minimized.
%%%%%%%%%%%%%%%%%%%%%%%%%%%%%%%%%%%%%%%
%%%%%%%%%%%%%%%%%%%%%%%%%%%%%%%%%%%%%%%
%\begin{figure*}[t]
%\begin{center}
%\includegraphics[width = \textwidth]{Immagini/Fig5.png}
%\caption{\label{fig:FIG5}{\ref{fig:FIG5}(a): Percentage error on the conformal map for the discretization of the continuous density distribution, as a function of the assigned number of radial elements for varying the number of tangential elements \ref{fig:FIG5}(b): Performance of the numerical model built discretizing the continuous density distribution, for different numbers of radial elements. \ref{fig:FIG5}(c): Discretized density distribution with $N_r=9$ and $N_\theta=17$.}}
%\end{center}
%\end{figure*}
%%%%%%%%%%%%%%%%%%%%%%%%%%%%%%%%%%%%%%%
%%%%%%%%%%%%%%%%%%%%%%%%%%%%%%%%%%%%%%%
Figure \ref{fig:FIG3}(d) shows the computed percentage error for various combinations of $N_\theta$ and $N_r$, showing that for each choice of $N_r$ there's an optimal $N_\theta$ value. We build numerical models of cloaks made by discretizing the theoretical continuous distribution in blocks and compute the scattered power, repeating the procedure for different $N_r$ values, in order to check the minimum number of radial blocks that allows for the computed figure of merit to converge to that of the continuous distribution. The results are shown in Figure \ref{fig:FIG3}(e) in terms of normalized scattered power over the uncloaked scenario, \editato{computed in the geometrical acoustic regime for the expected performance shown in Figure \ref{fig:FIG3}(c)}. We thus choose $N_r=9$ and the corresponding optimal $N_\theta=17$ as the best combination minimizing the overall number of blocks, while at the same time providing a sufficiently accurate discretization of the continuous density distribution. This in turns gives a total of 153 cells over one quarter of the ellipse to be optimized if one wants to "fill" each block with a unique rectangular cell. This is a considerably higher number of different cells to be optimized if compared to previously reported implementations for the axisymmetric cases, where only a few cell in radial direction need to be designed and then repeated without geometrical variations all over the azimuthal direction. Nonetheless, note that the radial variation of density is much higher than that in the tangential direction (Figure \ref{fig:FIG3}(f)), and this fact is thus exploited to reduce the computational burden associated to the high number of different optimization targets. 

Firstly, an optimization is performed for each radial "layer" targeting the cell showing the minimum density among those in the same layer. In this optimization the design variables are allowed to vary in a wide range. The stopping criterion of the algorithm is selected to be a maximum of 150 iterations or 40 consecutive generations without appreciable change in the cost function. The cost function is augmented in this case to consider also the density as:
\begin{equation}
    J=\frac{|C_{xx}-K_{ \theta}|}{K_{ \theta}}+\frac{|C_{yy}-K_{r}|}{K_{r}}+\frac{|\rho_{hom}- \rho_{target}|}{\rho_{target}}
\end{equation}
After this optimization, all the remaining cells are optimized, allowing for sensible changes only in the parameter $tr$ which to mostly affect density. All other parameters are constrained to take small variations around the values computed for the previous cell in the same layer. In this way convergence is always reached in less than 5 iterations. After the overall optimization process is carried out, figures of merit related to the closeness between obtained and expected material properties are computed and listed in Table \ref{tab:OptResults}.
\begin{table}[t]
	\centering
\begin{tabular}{c c c c c}
\centering
	 & target & $\mu$ & $\sigma$ & $\overline{e}_{\%}$\\
	\addlinespace[3pt]
	\hline
	\addlinespace[3pt]
	$K_{\theta }$ & $5.552\cdot 10^9$ Pa & $5.555\cdot 10^9$ Pa & $1.712\cdot 10^7$ Pa & 0.235\\
	$K_{r}$ & $0.9119\cdot10^9$ Pa & $0.9091\cdot 10^9$ Pa & $3.421\cdot10^6$ Pa & 0.416\\
	$K_{r \theta}$ & $2.25\cdot 10^9$ Pa & $2.209\cdot 10^9$ Pa & $9.593\cdot10^6$ Pa & 1.82\\
	$G$ & - & $7.052\cdot 10^7$ Pa & $4.754\cdot10^6$ Pa & -\\
	$\rho$ & - & - & - & 0.094\\
	\end{tabular}
	\caption{Summary of the results of the optimization with Genetic Algorithm.}
	\label{tab:OptResults}
\end{table}
For each element of the elasticity tensor, the mean value and the standard deviation calculated adopting the homogenized values of all the optimized unit cells are shown. The mean percentage error on $K_{r}$ and $K_{\theta}$ is well below $1\%$, while the error on $K_{r  \theta}$ is lower that 2$\%$, despite not having introduced it in the cost function. The shear modulus is two orders of magnitudes lower than the values of $K_{\theta}$. The mean percentage error on the scalar density value is the lowest and lies below 0.1$\%$. 
Once the unit cells are stacked in the prescribed order in cartesian coordinates, a simple linear map between the cartesian grid and the elliptical coordinate grid allows to accommodate the microstructure around the target in a conformal way \cite{quadrelli2021acoustic}, as mentioned in the previous.
%%%%%%%%%%%%%%%%%%%%%%%%%%%%%%%%%%%%%%%
%%%%%%%%%%%%%%%%%%%%%%%%%%%%%%%%%%%%%%%
\begin{figure*}[t!]
\begin{center}
\includegraphics[width = 0.85\textwidth]{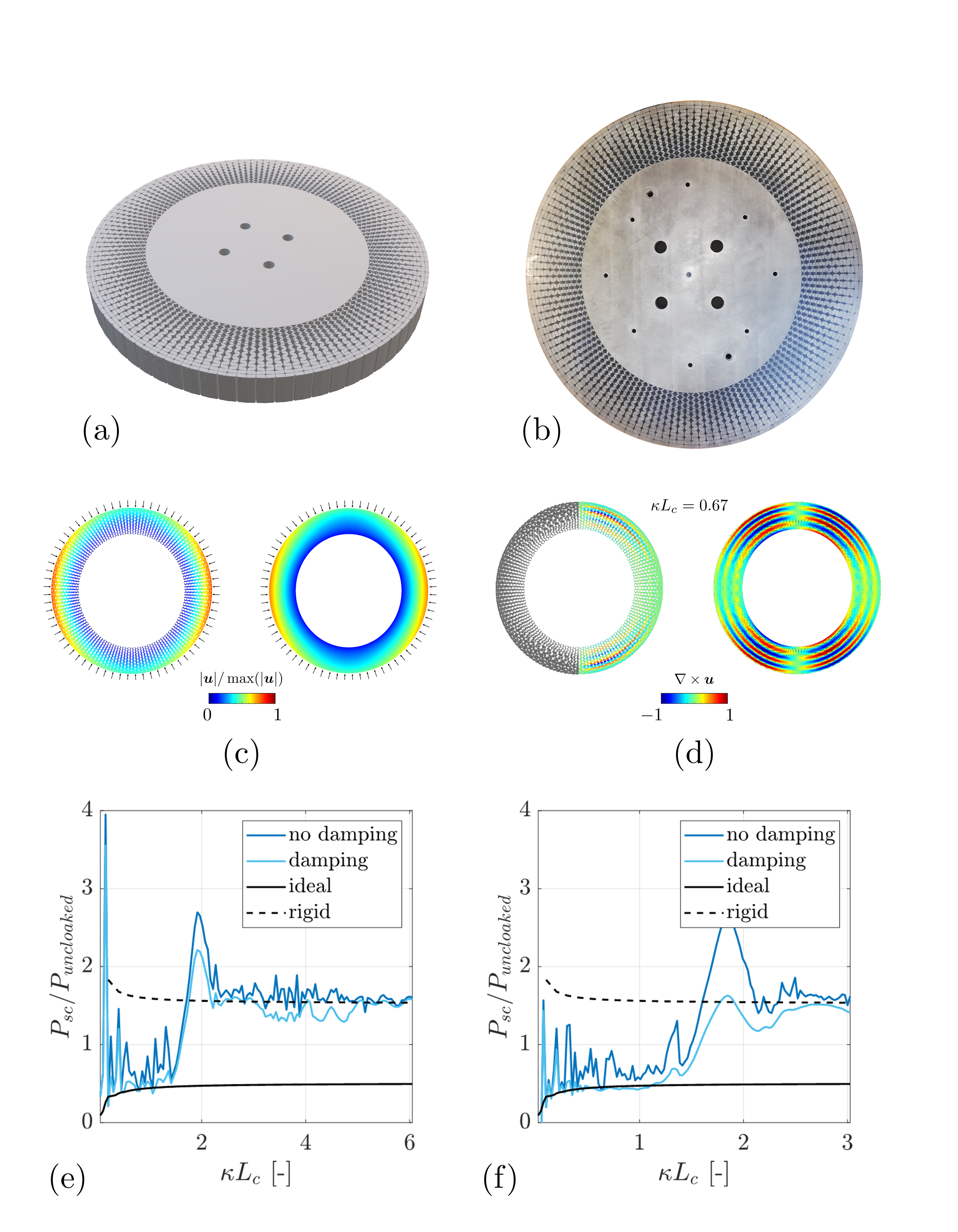}
\caption{\label{fig:FIG4}{\ref{fig:FIG4}(a): Schematic of the microstructured cloak. \ref{fig:FIG4}(b): Photo of the fabricated specimen. \ref{fig:FIG4}(c): Assessment of the homogenized elastic behavior of the microstructure. The displacements are compared for hydrostatic loading between the  microstructure and the corresponding homogenized material model. The inner surface of the cloak is fixed. \ref{fig:FIG4}(d): Shear dominated eigenmode of the cloak: the deformed shape and the curl of the displacement vector are shown and compared with those computed with the homogenized model. Black arrows show the direction of the displacement vectors. The curl is normalized with respect to the maximum computed for the homogenized case. \ref{fig:FIG4}(e): Numerical assessment of the scattered power normalized with respect to the uncloacked case, for vertical incidence. The black solid line refers to the scattering of an acoustically rigid obstacle shaped as $\partial \boldsymbol{\Omega}^-$, while the dashed one refers to $\partial \boldsymbol{\omega}^+$. Blue curve refers to numerical computations performed on the microstructure without considering structural damping, while light blue curve considers viscous damping ($\eta=0.01$). \ref{fig:FIG4}(f): Microstructured cloak built with 2x2 unit cells for each discretized block.}}
\end{center}
\end{figure*}
%%%%%%%%%%%%%%%%%%%%%%%%%%%%%%%%%%%%%%%
%%%%%%%%%%%%%%%%%%%%%%%%%%%%%%%%%%%%%%%
%%%%%%%%%%%%%%%%%%%%%%%%%%%%%%%%%%%%%%%
%%%%%%%%%%%%%%%%%%%%%%%%%%%%%%%%%%%%%%%
%\begin{figure*}[ht]
%\begin{center}
%\includegraphics[width = \textwidth]{Immagini/Fig7.png}
%\caption{\label{fig:FIG7}{Assessment of the homogenized elastic behavior of the microstructure. The displacements are compared for hydrostatic loading between the numerical model of the microstructure and the corresponding homogenized material model. The inner surface of the cloak is fixed.}}
%\end{center}
%\end{figure*}
%%%%%%%%%%%%%%%%%%%%%%%%%%%%%%%%%%%%%%%
%%%%%%%%%%%%%%%%%%%%%%%%%%%%%%%%%%%%%%%
%%%%%%%%%%%%%%%%%%%%%%%%%%%%%%%%%%%%%%%
%%%%%%%%%%%%%%%%%%%%%%%%%%%%%%%%%%%%%%%
%\begin{figure*}[ht]
%\begin{center}
%\includegraphics[width = \textwidth]{Immagini/Fig7bis.png}
%\caption{\label{fig:FIG7bis}{\editato{Shear dominated eigenmode of the cloak. The eigenmode computed with the microstructure are shown in terms of deformed shape and curl of the displacement vector, and compared with the ones computed with the homogenized model. Black arrows show the direction of the displacement vectors. The curl is normalized with respect to the maximum computed for the homogenized case.}}}
%\end{center}
%\end{figure*}
%%%%%%%%%%%%%%%%%%%%%%%%%%%%%%%%%%%%%%%
%%%%%%%%%%%%%%%%%%%%%%%%%%%%%%%%%%%%%%%
The obtained microstructure is shown in Figure \ref{fig:FIG4}(a). A common bandgap for the shear mode is computed to appear for all the unit cells comprised in the microstructure and lies between $ 1.79<\kappa L_c<2.46$, being $L_c$ the maximum size $L_\theta$ of the biggest unit cell in the microstructure once accommodated around the target. The normalized wavenumber is thus introduced as a measure of the relative size of the microstructure with respect to the wavelength in the background fluid.
Before proceeding with the numerical assessment of the performance of the microstructure, it is convenient to check whether homogenization based on the dispersion properties computed with periodic boundary conditions in cartesian coordinates still holds when different unit cells are joined together and deformed around an elliptic target. We thus perform a static analysis of the displacements of the microstructured cloak when hydrostatic pressure is applied outside it, with fixed boundary conditions on the inner surface, and compare the obtained displacement field with that of a cloak obtained assembling blocks characterized by the homogenized properties. The results are depicted for comparison in Figure \ref{fig:FIG4}(c) in terms of displacements normalized with respect to the maximum displacement of the microstructure: they show good agreement, demonstrating that the homogenized cloak correctly represents the elastic properties of the assembled microstructure. To check whether density is also correctly reproduced, the first set of eigenmodes of the microstructure and that of a cloak made with the homogenized materials are computed and compared, showing good agreement (the interested reader can find a comparison in the \ref{App1}).
\editato{Since the shear modulus is different from zero, shear modes are also observed, as depicted in Figure \ref{fig:FIG4}(d).}
Finally, fully coupled acoustic structural simulations are conducted to assess the scattering performance of the microstructure against plane wave incidence. The results are shown in Figure \ref{fig:FIG4}(e), where the normalized scattered power is shown for vertical incidence, comparing the ideal results with the computed performance of the microstructure. The blue line refers to the simulation without damping, while the light blue refers to simulations performed considering the structural viscous damping of Aluminum, which is included in the model via a complex Young's modulus:
\begin{equation}
    E=E_0(1-j\eta)
\end{equation}
with $\eta=0.01$. \editato{It can be seen that below a certain $\kappa L_c$ threshold where the quasi-static homogenization holds, the scattered power of the microstructure is close to that expected. The observed peaks witness the presence of resonances related to the non null shear modulus. Indeed, since the longitudinal modes are impedance matched with the surrounding fluid on the outer surface of the cloak only the shear waves can lead to cavity resonances. Damping is beneficial in this case to mitigate the effect of those resonances. For high values of $\kappa L_c$ the scattered power tends to the value that is obtained with an acoustically rigid surface with the same shape of the outer surface of the cloak. This results from the homogenization not holding anymore, and the geometrical acoustic limit. In between the two regimes, no resonances are observed, but a big peak in the scattered power. This regime corresponds to the $\kappa L_c$ values of the pentamode bandgap. The single-mode bandgap has been considered the ideal working frequency of pentamode materials, since no bulk propagating shear waves can propagate in the microstructure, which thus behaves as a fluid. The lack of propagative shear modes explains indeed the absence of resonances. However, our numerical results show that the pentamode bandgap seems to witness the worst performance in terms of scattering cross section. This can be  explained by considering that while the bulk propagation properties of a microstructure can be inspected by checking the propagative modes alone along with the location of the bandgaps, the interface phenomena, as for example the transmission and reflection across a fluid/microstructure interface, depend on the interaction between the incoming wave and the full set of Bloch modes, evanescent ones included, unless they are deaf \cite{laude2011bloch}. While a detailed study of the interaction between fluids and pentamode materials is not the goal of this paper, it seems that the presence of the pentamode bandgap enhances a mirror-like behavior of the system, thus deteriorating performace. }  A rule of thumb can be deduced for the cut-off frequency that sets the working limit of the microstructure, by building another cloak characterized by a 2x2 pattern of equal unit cells inside each block of the discretized material parameters. In this way, the homogenized cloak is the same, but the size of the unit cells (and thus $L_c$) is reduced to one half. Performing on this newly assembled numerical model the computation of the scattered power, the observed dependence on $\kappa L_c$ is maintained, as shown in Figure \ref{fig:FIG4}(f).
%%%%%%%%%%%%%%%%%%%%%%%%%%%%%%%%%%%%%%%%%%%%%%%%%%%%%%%%%%%%%
%%%%%%%%%%%%%%%%%%%%%%%%%%%%%%%%%%%%%%%%%%%%%%%%%%%%%%%%%%%%%
\section{Experimental setup}
%%%%%%%%%%%%%%%%%%%%%%%%%%%%%%%%%%%%%%%%%%%%%%%%%%%%%%%%%%%%%
%%%%%%%%%%%%%%%%%%%%%%%%%%%%%%%%%%%%%%%%%%%%%%%%%%%%%%%%%%%%%
%%%%%%%%%%%%%%%%%%%%%%%%%%%%%%%%%%%%%%%
%%%%%%%%%%%%%%%%%%%%%%%%%%%%%%%%%%%%%%%
%\begin{figure*}[ht]
%\begin{center}
%\includegraphics[width = 0.8\textwidth]{Immagini/Fig9.png}
%\caption{\label{fig:FIG9}{Schematic of the measurement system.}}
%\end{center}
%\end{figure*}
%%%%%%%%%%%%%%%%%%%%%%%%%%%%%%%%%%%%%%%
%%%%%%%%%%%%%%%%%%%%%%%%%%%%%%%%%%%%%%%
\editato{A specimen is produced to be acoustically probed by an impinging signal with spectrum corresponding to $0.45<\kappa L_c < 1.34$, i.e. the numerically computed working range.} The microstructure designed following the method detailed in the previous section is manufactured by wire EDM from an Aluminum plate, resulting in the component shown in Figure \ref{fig:FIG4}(b). The cloak is attached to an Aluminum elliptic obstacle, to implement a rigid boundary condition inside the inner surface of the cloak. To create a bidimensional acoustic environment, the specimen is placed in a waveguide made by two Aluminum plates \editato{sized $10.5\lambda_s \times 10.5\lambda_s\times 0.05\lambda_s$, being $\lambda_s$ the wavelength at the central frequency of the signal}. The pressure compensation technique previously illustrated in \cite{zheng2018two} is adopted, i.e. an aluminum elliptical disk shaped as $\partial \boldsymbol{\Om}^-$ is placed outside each Aluminum plate in correspondence with the specimen. The obtained "sandwich" is suspended vertically in the tank in such a way that the cloak is equally distant from the bottom and the free surface of the pool. An omnidirectional projector (co.l.mar.  TD0720) is placed $32\lambda_s$ away from the center of the sandwich, in such a way that the pressure field at the aperture of the waveguide is as constant as possible along the spacing between the plates, due to spherical spreading. The projector is driven by a KEYSIGHT 33500B signal generator amplified by a REVAC Pro 1200 model AR 446. The acoustic field inside the waveguide is reconstructed adopting a hydrophone (co.l.mar. TD0190) that is moved in successive positions to scan the area around the specimen via a gantry robot. The spacial distancing between measuring points is selected to be \editato{$0.2\lambda_s$}. The signal of the hydrophone is pre-amplified with a B\&K type 2650 amp and then acquired with a NI-9222 board for post processing. Measurements are also conducted along three circumferences of radius \editato{3.84$\lambda_s$, 3.89$\lambda_s$ and 3.94$\lambda_s$ } centered around the specimen, with an angular discretization of 5 degrees. These measurements will be used to compute the polar dependence of scattered intensity. A set of measurements is conducted for the specimen placed with the major axis parallel to the propagation direction (vertical incidence, best case scenario), afterwards the sandwich is rotated by 90 degrees and measurements are performed with the major axis perpendicular to the propagation direction (horizontal incidence, worst case scenario). \editato{To prevent flooding, the holes are sealed with a layer of EPDM rubber in between the specimen and the plates implementing the waveguide}.  The same setup is used for measurements on an elliptical Aluminum cylinder representing the uncloaked target, and for a smaller Aluminum cylinder shaped as $\partial \boldsymbol{\Om}^-$, for critical comparison of results. \editato{The EPDM rubber is used also for this latter cases, in order to have as equal experimental conditions as possible with respect to the cloaked case.}
%%%%%%%%%%%%%%%%%%%%%%%%%%%%%%%%%%%%%%%%%%%%%%%%%%%%%%%%%
%%%%%%%%%%%%%%%%%%%%%%%%%%%%%%%%%%%%%%%%%%%%%%%%%%%%%%%%%
\section{Results and Discussion}
%%%%%%%%%%%%%%%%%%%%%%%%%%%%%%%%%%%%%%%%%%%%%%%%%%%%%%%%%
%%%%%%%%%%%%%%%%%%%%%%%%%%%%%%%%%%%%%%%%%%%%%%%%%%%%%%%%%
%%%%%%%%%%%%%%%%%%%%%%%%%%%%%%%%%%%%%%%
%%%%%%%%%%%%%%%%%%%%%%%%%%%%%%%%%%%%%%%
\begin{figure*}[ht!]
\begin{center}
\includegraphics[width = 0.8\textwidth]{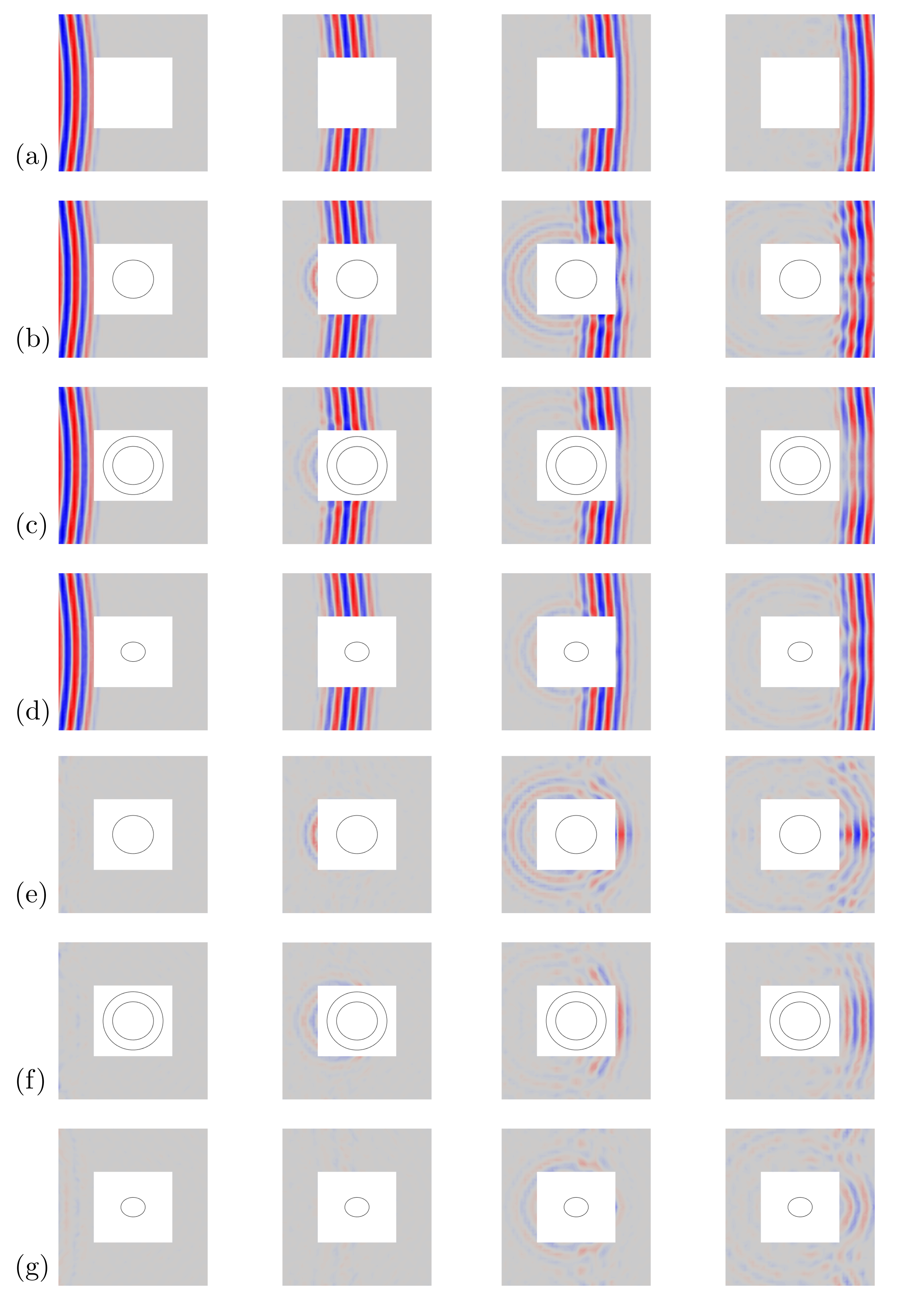}
\caption{\label{fig:FIG5}{Measured acoustic fields. \ref{fig:FIG5}(a): four snapshot of the incident wave traveling in empty waveguide. \ref{fig:FIG5}(b): Same four snapshot as in \ref{fig:FIG5}(a) but measured with the uncloaked obstacle in the waveguide. \ref{fig:FIG5}(c): Cloaked obstacle. \ref{fig:FIG5}(d): Smaller equivalent obstacle shaped as $\partial \boldsymbol{\Omega}^-$. \ref{fig:FIG5}(e): Scattered field for the uncloaked obstacle computed as the difference between the measured signal and the signal measured with the empty waveguide shown in \ref{fig:FIG5}(a). \ref{fig:FIG5}(f): Cloaked obstacle. \ref{fig:FIG5}(g): Smaller equivalent obstacle shaped as $\partial \boldsymbol{\Omega}^-$.}}
\end{center}
\end{figure*}
%%%%%%%%%%%%%%%%%%%%%%%%%%%%%%%%%%%%%%%
%%%%%%%%%%%%%%%%%%%%%%%%%%%%%%%%%%%%%%%
The time histories measured in subsequent adjacent points inside the waveguide are used to reconstruct the acoustic field that is shown in Figure \ref{fig:FIG5}. In particular, Figure \ref{fig:FIG5}(a) represents four different time instants of the propagation of the signal from left to right inside the empty waveguide, thus showing the incident burst. Figure \ref{fig:FIG5}(b) shows instead the same four snapshots, but measured with the uncloaked Aluminum obstacle. In this case the scattered fields can be clearly seen in the two panels in the middle of Figure \ref{fig:FIG5}(b) where a circular backscattering wavefront is observed. In the panel to the right, the distortion of the wavefronts after the obstacle witness the forward scattering. The acoustic field at the same four instants is also plotted in Figure \ref{fig:FIG5}(c) and \ref{fig:FIG5}(d) respectively for the cloaked case, and for the Aluminum obstacle shaped as $\partial \boldsymbol{\Omega}^-$, which thus represents the reference expected behavior of the cloak. Full animated gifs of the measured acoustic fields can be found in the Supplementary Material, both for the vertical and horizontal incidence cases. Comparing Figure \ref{fig:FIG5}(b) with Figure \ref{fig:FIG5}(c), it can be clearly appreciated the reduction in backscattering that is observed in the cloaked case, and also the reduced distortion of the wavefronts after the cloaked obstacle. Nonetheless, it can be also observed the reduction in intensity of the field in the forward direction, which witnesses absorption inside the cloak. \editato{Moreover, backscattering from the cloak should be expected to occur at the same time as occurs for the equivalent target, which is instead delayed due to the reduced size.} The backscattering that is observed in the cloaked case as soon as the wave impinges on the cloak thus results from a small impedance mismatch between the outer surface of the cloak and the background fluid, which is reasonable since standard material parameters have been considered in the design phase for water ($\rho_w=1000\;[kg/m^3]$ and $c_w=1500\;[m/s]$) that can differ from the actual experimental values. The same phenomenology can be observed also for the measured fields in case the obstacles are rotated 90 degrees with respect to the incidence direction. 
%%%%%%%%%%%%%%%%%%%%%%%%%%%%%%%%%%%%%%%
%%%%%%%%%%%%%%%%%%%%%%%%%%%%%%%%%%%%%%%
%\begin{figure*}[ht]
%\begin{center}
%\includegraphics[width = 0.8\textwidth]{Immagini/Fig11.png}
%\caption{\label{fig:FIG11}{\ref{fig:FIG11}(a): four snapshot of the incident wave traveling in empty waveguide. \ref{fig:FIG11}(b): Scattered field for the uncloaked obstacle computed as the difference between the measured signal and the signal measured with the empty waveguide shown in \ref{fig:FIG11}(a). \ref{fig:FIG11}(c): Cloaked obstacle. \ref{fig:FIG11}(d): Smaller equivalent obstacle shaped as $\partial \boldsymbol{\Omega}^-$.}}
%\end{center}
%\end{figure*}
%%%%%%%%%%%%%%%%%%%%%%%%%%%%%%%%%%%%%%%
%%%%%%%%%%%%%%%%%%%%%%%%%%%%%%%%%%%%%%%
Figure \ref{fig:FIG5}(e)-(f)-(g) show the scattered fields computed subtracting the incident field (Figure \ref{fig:FIG5}(a)) to the measured total fields represented in \ref{fig:FIG5}(b)-(c)-(d). The computation of the scattered field allows for evaluation of the scattered intensity as a function of the polar angle. This is done adopting the measurements conducted around three circumferences having increasing radius, together with the definition of instantaneous intensity in radial direction:
\begin{equation}
    I_r=pv_r
\end{equation}
The pressure adopted is the one measured at the circumference with mean radius, while the measurements taken on the external and internal circumferences are used to compute the radial velocity with the Euler formula
\begin{equation}
    v_r=-\frac{1}{\rho_w}\int \frac{\partial p}{\partial r}dt
\end{equation}
where the radial gradient of pressure has been approximated with central differences. The error introduced by such approximation can be computed as \cite{rossing2007springer}:
\begin{equation}
    e_\%= 1-\frac{\sin(\kappa \Delta r)}{\kappa \Delta r}
\end{equation}
$\delta r$ being the discrete spacing between the circumferences, and $\kappa$ the wavenumber. This error is computed to be $\approx 2 \%$ at the central frequency of the considered signal. The instantaneous intensity at each polar direction is then used to compute the mean intensity by time averaging.
%%%%%%%%%%%%%%%%%%%%%%%%%%%%%%%%%%%%%%%
%%%%%%%%%%%%%%%%%%%%%%%%%%%%%%%%%%%%%%%
\begin{figure*}[ht]
\begin{center}
\includegraphics[width = 0.8\textwidth]{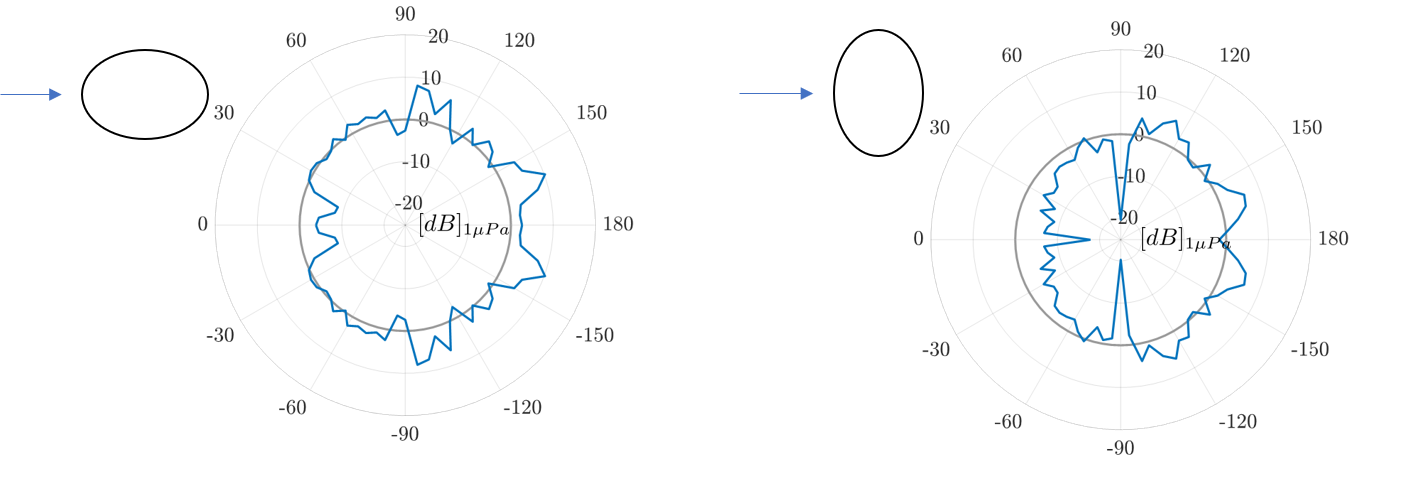}
\caption{\label{fig:FIG12}{\editato{Polar plots of the difference between the experimental and numerical Decibel gains of scattered acoustic intensity, for both the measured directions of incidence.}}}
\end{center}
\end{figure*}
%%%%%%%%%%%%%%%%%%%%%%%%%%%%%%%%%%%%%%%
%%%%%%%%%%%%%%%%%%%%%%%%%%%%%%%%%%%%%%%
\editato{This allows in turn to compute the Decibel gain in scattered intensity as a difference between the cloaked case and the uncloaked scenario. This gain is then compared to that obtained averaging over the same frequency range the results of the numerical simulations, by computing the difference between the experimental and numerical one. The result is shown in Figure \ref{fig:FIG12}. It is thus understood that the performance of the cloak are better than those expected in the backscattering direction, while are worse than expected in the forward scattering. This clearly witnesses the presence of some absorption mechanism greater than that was accounted in the numerical prediction, that absorbs without reflection part of the incident radiation, leaving a deeper shadow behind the obstacle.} 
%%%%%%%%%%%%%%%%%%%%%%%%%%%%%%%%%%%%%%%%%%%%%%%%%%%%%%%%
%%%%%%%%%%%%%%%%%%%%%%%%%%%%%%%%%%%%%%%%%%%%%%%%%%%%%%%%
\section{Conclusion}
%%%%%%%%%%%%%%%%%%%%%%%%%%%%%%%%%%%%%%%%%%%%%%%%%%%%%%%%
%%%%%%%%%%%%%%%%%%%%%%%%%%%%%%%%%%%%%%%%%%%%%%%%%%%%%%%%
In this manuscript a cloak for elliptically shaped obstacles aiming at reducing the scattered acoustic power has been designed, fabricated and experimentally validated, providing the first experimental design of an underwater acoustic cloak for non axisymmetric objects. The challenges implied by the lack of axial symmetry of the problem has been tackled by setting a transformation that allows for the material property inhomogeneity to depend only on the density distribution inside the cloak, leaving the anisotropic stiffness constant. This allowed great simplification of the optimization problem and of the assembly of the microstructure, that has been conducted in a conformal way exploiting the properties of the elliptic coordinate system. The designed microstructure is numerically simulated to gain insight in the cutoff frequency above which quasi-static homogenization does not hold. It is observed that as a rule of thumb, frequency having wavelength above four time the size of the biggest cells define the working frequency range of the microstructure. A common pentamode bandgap was also numerically computed to occur for all the different unit cells in the microstructure. Despite causing no resonances to occur, the appearance of such bandgap seems to decrease the performance of the system in terms of scattering cross section. Shear mode resonances are instead responsible of the peaks in scattered power observed below the bandgap. These are actually mitigated by the structural damping in the microstructure which thus play a crucial role for obtaining good cloaking results. Absorption is also observed to play a fundamental role in the actual experimental measured performance, which are better than expected in backward scattering, but worse in forward scattering, due to the absorbed acoustical power. The absorption mechanism that is responsible of this variation of performance is not yet fully understood, since seems to be too high to be attributed to the viscosity of aluminum only. That could be related to dissipation induced by resonances and viscous losses of the air in the tiny cavities of each unit cell. A more in depth analysis of this phenomenon is left for future work, and could shed light on the necessity of considering the air instead of vacuum in the analysis of the performance of pentamode materials for acoustic applications.

\section{Acknowledgements}
The authors are grateful to the Italian Ministry of Defense for the financial support granted to this work through the PNRM "SUWIMM". NGM s.r.l. is acknowledged for the fabrication of the experimental device. The authors are also thankful to co.l.mar. s.r.l., Prof. Davide Tarsitano and Giacomo Ferrari for the help given during the experimental campaign.

\appendix
\section{Mode Shapes of the Elliptical Cloak}
\label{App1}
%%%%%%%%%%%%%%%%%%%%%%%%%%%%%%%%%%%%%%%
%%%%%%%%%%%%%%%%%%%%%%%%%%%%%%%%%%%%%%%
\begin{figure*}[ht]
\begin{center}
\includegraphics[width = 0.8\textwidth]{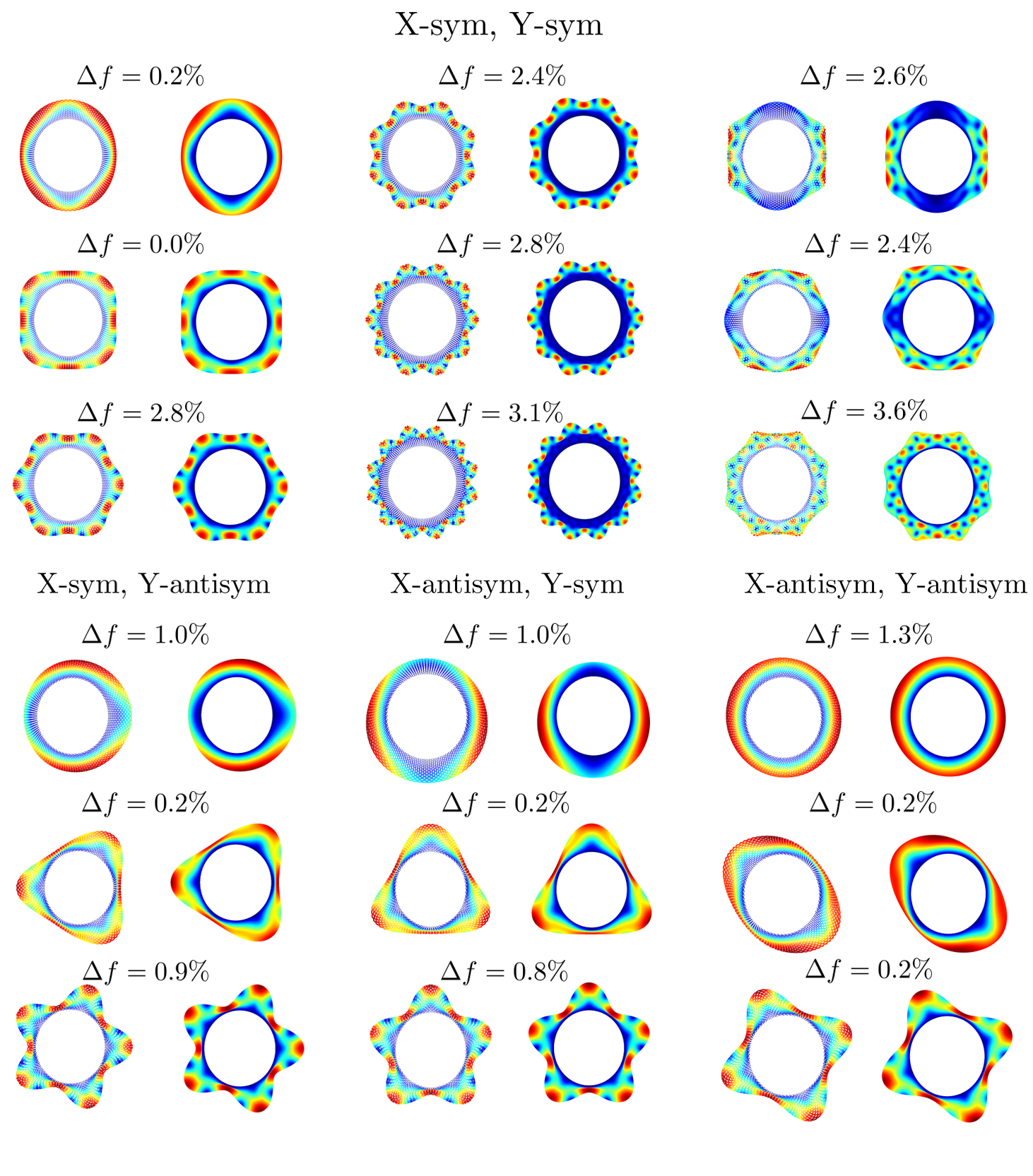}
\caption{\label{fig:App1}{Mode shapes of the cloak computed on the actual microstructure and on the homogenized model. The percentage difference in natural frequency is also reported.}}
\end{center}
\end{figure*}
\begin{figure*}[ht]
\begin{center}
\includegraphics[width = 0.8\textwidth]{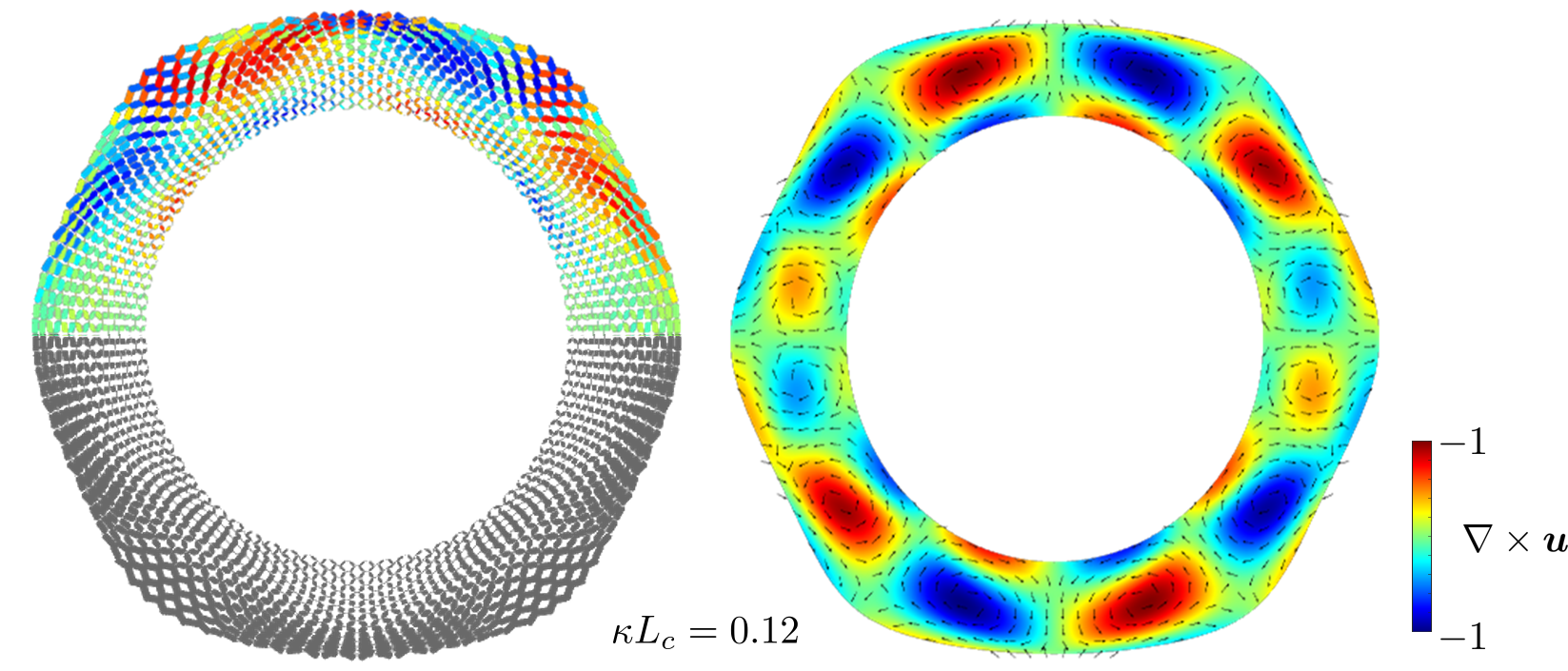}
\caption{\label{fig:App2}{Mode shape and curl for a the mode at $\kappa L_c=0.12$, that corresponds to the second in the third column of Figure \ref{fig:App1}.}}
\end{center}
\end{figure*}

%%%%%%%%%%%%%%%%%%%%%%%%%%%%%%%%%%%%%%%
%%%%%%%%%%%%%%%%%%%%%%%%%%%%%%%%%%%%%%%
In this part the results of the numerical calculation of the eigenfrequencies and modes of vibration of the assembled microstructured cloak is reported. In Figure \ref{fig:App1} the mode shapes are shown and compared between those obtained by direct calculation on the microstructure, and those obtained using the homogenized model. Along with each pair of mode shapes, the percentage difference in eigenfrequency is reported. Since the system has two axis of symmetry, the modes are split based on the degree of symmetry/antisymmetry shown by the associated displacement field. The good agreement between the behavior of the microstructure and that of the model built with the homogenized properties assures about the fact that the quasi-static homogenization computed in cartesian coordinates still holds even if only one unit cell of each different type is used and even if each cell is slightly deformed during accommodation around the target. In particular, since the elastic properties can be checked with the simple static analysis as the one shown in Figure \ref{fig:FIG4}(c), the assessment via eigenmodes is useful to assess the equivalence of the inertial properties. Among the computed eigenmodes, they can be recognized the whispering gallery modes where stationary waves form on the surface of the cloak when the perimeter is a multiple of half the wavelength, like the ones depicted in the first two columns of the modes sym-sym modes. The three in the last column of the sym-sym modes also arise from waves running back and forth in the tangential direction,  but are instead dominated by local rotation of the bulk of the cloak, as is shown in Figure \ref{fig:App2}.
%%%%%%%%%%%%%%%%%%%%%%%%%%%%%%%%%%%%%%%
%%%%%%%%%%%%%%%%%%%%%%%%%%%%%%%%%%%%%%%
%%%%%%%%%%%%%%%%%%%%%%%%%%%%%%%%%%%%%%%
%%%%%%%%%%%%%%%%%%%%%%%%%%%%%%%%%%%%%%%
When the thickness of the cloak is a multiple of half the wavelength of shear waves, modes like the one depicted in Figure \ref{fig:FIG4}(d) are instead observed.
\bibliographystyle{elsarticle-num}
\bibliography{Mybibliography}

\begin{thebibliography}{10}
\expandafter\ifx\csname url\endcsname\relax
  \def\url#1{\texttt{#1}}\fi
\expandafter\ifx\csname urlprefix\endcsname\relax\def\urlprefix{URL }\fi
\expandafter\ifx\csname href\endcsname\relax
  \def\href#1#2{#2} \def\path#1{#1}\fi

\bibitem{chen2010acoustic}
H.~Chen, C.~T. Chan, Acoustic cloaking and transformation acoustics, Journal of
  Physics D: Applied Physics 43~(11) (2010) 113001.

\bibitem{Pendry2006}
J.~B. Pendry, D.~Schurig, D.~R. Smith, Controlling electromagnetic fields,
  Science 312 (2006) 1780--1782.

\bibitem{Leonhardt2006}
U.~Leonhardt, Optical conformal mapping, Science 312 (2006) 1777--1780.

\bibitem{Norris2008}
A.~N. Norris, Acoustic cloaking theory, Proceedings of the Royal Society A 464
  (2008) 2411--2434.

\bibitem{cummer2007one}
S.~A. Cummer, D.~Schurig, One path to acoustic cloaking, New Journal of Physics
  9~(3) (2007) 45.

\bibitem{chen2007acoustic}
H.~Chen, C.~Chan, Acoustic cloaking in three dimensions using acoustic
  metamaterials, Applied physics letters 91~(18) (2007) 183518.

\bibitem{torrent2008acoustic}
D.~Torrent, J.~S{\'a}nchez-Dehesa, Acoustic cloaking in two dimensions: a
  feasible approach, New Journal of Physics 10~(6) (2008) 063015.

\bibitem{zhang2011broadband}
S.~Zhang, C.~Xia, N.~Fang, Broadband acoustic cloak for ultrasound waves,
  Physical review letters 106~(2) (2011) 024301.

\bibitem{milton1995elasticity}
G.~W. Milton, A.~V. Cherkaev, Which elasticity tensors are realizable? (1995).

\bibitem{chen2015latticed}
Y.~Chen, X.~Liu, G.~Hu, Latticed pentamode acoustic cloak, Scientific reports 5
  (2015) 15745.

\bibitem{chen2017broadband}
Y.~Chen, M.~Zheng, X.~Liu, Y.~Bi, Z.~Sun, P.~Xiang, J.~Yang, G.~Hu, Broadband
  solid cloak for underwater acoustics, Physical Review B 95~(18) (2017)
  180104.

\bibitem{li2012homogeneous}
T.~Li, M.~Huang, J.~Yang, Y.~Lan, J.~Sun, Homogeneous material constructed
  acoustic cloak based on coordinate transformation, Journal of vibration and
  acoustics 134~(5) (2012).

\bibitem{li2014two}
Q.~Li, J.~S. Vipperman, Two-dimensional acoustic cloaks of arbitrary shape with
  layered structure based on transformation acoustics, Applied Physics Letters
  105~(10) (2014) 101906.

\bibitem{li2018non}
Q.~Li, J.~S. Vipperman, Non-singular three-dimensional arbitrarily shaped
  acoustic cloaks composed of homogeneous parts, Journal of Applied Physics
  124~(3) (2018) 035103.

\bibitem{li2019two}
Q.~Li, J.~S. Vipperman, Two-dimensional arbitrarily shaped acoustic cloaks with
  triangular patterns of homogeneous properties, Journal of Vibration and
  Acoustics 141~(2) (2019).

\bibitem{chen2016design}
Y.~Chen, X.~Liu, G.~Hu, Design of arbitrary shaped pentamode acoustic cloak
  based on quasi-symmetric mapping gradient algorithm, The Journal of the
  Acoustical Society of America 140~(5) (2016) EL405--EL409.

\bibitem{quadrelli2021acoustic}
D.~E. Quadrelli, G.~Cazzulani, S.~La~Riviera, F.~Braghin, Acoustic scattering
  reduction of elliptical targets via pentamode near-cloaking based on
  transformation acoustics in elliptic coordinates, arXiv preprint
  arXiv:2105.08695 (2021).

\bibitem{laude2011bloch}
V.~Laude, R.~P. Moiseyenko, S.~Benchabane, N.~F. Declercq, Bloch wave deafness
  and modal conversion at a phononic crystal boundary, AIP Advances 1~(4)
  (2011) 041402.

\bibitem{zheng2018two}
M.~Zheng, Y.~Chen, X.~Liu, G.~Hu, Two-dimensional water acoustic waveguide
  based on pressure compensation method, Review of Scientific Instruments
  89~(2) (2018) 024902.

\bibitem{rossing2007springer}
T.~D. Rossing, T.~D. Rossing, Springer handbook of acoustics, Vol.~1, Springer,
  2007.

\end{thebibliography}

\end{document}